\documentclass[preprint]{emulateapj}

\slugcomment{Submitted to the Astrophysical Journal Letters}
\shortauthors{Zaritsky, Gil de Paz, \& Bouquin}
\shorttitle{}

\begin{document}
\title{An Empirical Connection between the UV Color of Early Type Galaxies and the Stellar Initial Mass Function}
  
\author{Dennis Zaritsky\altaffilmark{1}, Armando Gil de Paz\altaffilmark{2}, and Alexandre Y. K. Bouquin\altaffilmark{2}}
\altaffiltext{1}{Steward Observatory, University of Arizona, 933 North Cherry Avenue, Tucson, AZ 85721, USA; dennis.zaritsky@gmail.com}
\altaffiltext{2}{Departamento de Astrof\'{\i}sica y CC$.$ de la Atm\'osfera, Facultad de CC$.$ F\'{\i}sicas, Universidad Complutense de Madrid, Avda$.$ de la Complutense s/n, Madrid E-28040, Spain}


\begin{abstract} 
Using new UV magnitudes for a sample of early-type galaxies, ETGs, with published stellar mass-to-light ratios, $\Upsilon_*$'s, we find a correlation between UV color and $\Upsilon_*$  that is tighter than those previously identified between $\Upsilon_*$ and either the central stellar velocity dispersion, metallicity, or alpha enhancement. The sense of the correlation is that galaxies with larger $\Upsilon_*$ are bluer in the UV.  We conjecture that differences in the lower mass end of the stellar initial mass function, IMF, are related to the nature of the extreme horizontal branch populations that are generally responsible for the UV flux in ETGs. If so, then UV color can be used to identify ETGs with particular IMF properties and to estimate $\Upsilon_*$.
\end{abstract}

\keywords{stars: luminosity function, mass function --- galaxies:elliptical and lenticular, cD --- galaxies: stellar content --- galaxies: evolution}

\section{Introduction}
\label{sec:intro}

We present and discuss data that are integral to two key unresolved questions regarding the stellar populations of early type galaxies. Do variations exist among the stellar initial mass functions of early-type galaxies (ETGs)? What is the physical mechanism for producing the stars that give rise to the UV flux in ETGs?

A spate of recent results suggest that the stellar initial mass function (IMF) of giant ETGs has more low mass stars for a given total stellar mass than predicted even by the relatively extreme Salpeter mass function ($dN/dM \propto M^{-2.35}$). Specifically, \cite{vandokkum}, \cite{spiniello}, and \cite{ferreras} reach such a conclusion using spectral line indices that are sensitive to the dwarf-to-giant ratio. Independently, \cite{cappellari} and \cite{laesker} find consistent results using dynamical models to measure the total mass.
These results are to be contrasted with the extensive evidence in the local neighborhood for an IMF that turns over at sub-solar stellar masses \citep{bastian}.

Although the evidence for variations in the bottom portion of the IMF now extends beyond studies of ETGs, with stellar clusters providing some of the most direct evidence \citep{strader,z12,z13}, and direct imaging that resolves sub-solar mass
stellar populations promising eventually to settle the matter \citep{kalirai}, the analysis of ETGs is critical in our efforts to understand galaxy evolution. 

The second of our two questions
traces its origin to the observation of an unexpectedly large UV flux from ETGs \citep{code}. Explanations proposed for this ``UV-excess" or ``UV-upturn" fall into two classes ever since the discovery paper: young stars \citep{gunn,rv} and hot evolved stars.  Although in some cases there may be a connection to recent, residual star formation \citep{yi,boselli}, this explanation does not account for the majority of the UV-upturns \citep{ferguson,brown97,brown00,boselli,han} even though it plays a more prominent role in intermediate and low mass ETGs \citep{boselli,kannappan} and in lenticulars \citep{salim}. 

Various scenarios involving evolved stars have been put forward including, post-asymptotic giant branch stars \citep{rose}, hot horizontal branch stars \citep{ciardullo,bruzual}, and accreting white dwarfs \citep{greggio}. Filling in the details of these populations has proved difficult, but a preliminary consensus is that hot, or extreme, horizontal branch (EHB) stars must be the key contributor to the UV fluxes of ETGs \citep{han,lisker}. Although models depend on numerous poorly constrained parameters (mass ratio distribution, tidal atmospheric stripping efficiency), they manage to reproduce several key observational characteristics of the UV-upturn population and have the benefit of relying on a population of objects that are directly observed to exist \citep{brown00,lisker,buzzoni12}. 

The relevance of the UV-upturn extends beyond the nature of EHB stars. For example, models of EHB evolution may inform how the AGB is populated, which will affect how to compute the mass-to-light ratio in bands where the AGB contribution is significant \citep[see][]{buzzoni}. In general, 
investigators are focusing on the evolution of binary systems \citep{han,lisker} as a path to the formation of EHB stars. As expected, some of these models do predict a connection between UV excesses and AGB properties \citep{buzzoni}, thereby implying an effect on calculations of the mass-to-light ratio. Conversely, variations in the IMF could directly affect the UV-upturn population, particularly if binary systems with large mass ratios are an important progenitor class for EHBs. Our two questions are, therefore, intricately connected.

UV-upturn properties vary widely among galaxies \citep[see][]{code,burstein,gdp,bureau}. Among the parameters that have been related to the strength of the UV-upturn are 1) stellar population age, which is particularly relevant if some fraction of the UV-bright population is relatively young \citep{yi,kav,scha}, 2) galaxy mass, which is relevant if the galaxy's evolutionary history and total stellar mass are tied to each other \citep{boselli}, 3) stellar density, which is relevant if density-dependent binary processes play a role in forming EHB stars and densities are sufficiently large \citep[see][for evidence of such a relationship in globular clusters]{peacock}, and 4) metallicity, which is relevant because chemical abundance plays a role in stellar evolution, particularly in late-stage evolution where mass loss is significant. Interestingly, a promising correlation was found between UV properties and  metallicity, with more prominent upturns seen in galaxies with greater Mg$_2$ Lick index \citep{faber,burstein}, an index which broadly tracks metallicity even though it does have some sensitivity to gravity. It is critical to continue the search for other patterns and investigate all such possibilities as a source of insight into the physical mechanisms at work to tie all of this together into a coherent framework.

We examine a set of well-studied galaxies for which line-index based stellar mass-to-light ratios, $\Upsilon_*$, are already available, as are metallicity diagnostics, internal velocity dispersions, and absolute luminosities \citep[from][and references therein]{conroy}. We present new measurements of the FUV and NUV fluxes from $GALEX$ data for this set of galaxies. In \S2 we discuss the data and in \S3 we describe the findings of our study.

\section{The Data and Measurements}
\label{sec:data}

The parent sample for this study is that of nearby ETGs studied by \cite{conroy}, which overlaps extensively with the sample defined by the SAURON team \citep{bureau,jeong}. 
Our addition to the existing data for this sample is the set of homogeneous measurements of the FUV (1350---1750\AA) and NUV (1750---2750\AA) photometry made possible by the Galaxy Evolution Explorer ($GALEX$) satellite \citep{martin}.

\begin{figure}
\plotone{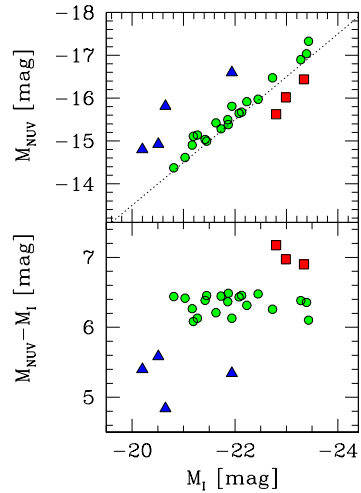}
\caption{UV properties of sample galaxies. Galaxies noted with blue circles or red squares are outliers described in text.}
\label{fig:UV}
\end{figure} 

\begin{deluxetable}{lrrr}
\tablecaption{Galaxy Sample and Properties\tablenotemark{a}}
\tablewidth{0pt}
\tablehead{
\colhead{Name}&
\colhead{$m_{FUV}$}&
\colhead{$m_{NUV}$}&
\colhead{Distance}\\
&&&[Mpc]\\
}
\startdata
NGC474&18.32$\pm$0.10&15.93$\pm$0.06&32.0\\
NGC524&17.76$\pm$0.04&15.82$\pm$0.04&23.3\\
NGC821&18.43$\pm$0.14&16.23$\pm$0.09&23.4\\
NGC1023&16.54$\pm$0.08&14.86$\pm$0.06&11.1\\
NGC2549&18.38$\pm$0.05&16.47$\pm$0.03&12.3\\
NGC2685&15.53$\pm$0.07&15.07$\pm$0.02&15.0\\
NGC2695&18.44$\pm$0.05&17.07$\pm$0.06&31.5\\
NGC2699&19.42$\pm$0.04&17.72$\pm$0.04&26.2\\
NGC2768&17.05$\pm$0.06&15.22$\pm$0.04&21.8\\
NGC2974&17.37$\pm$0.56&15.79$\pm$0.26&20.9\\
NGC3377&17.18$\pm$0.27&15.08$\pm$0.02&10.9\\
NGC3379&16.35$\pm$0.01&14.82$\pm$0.02&10.3\\
NGC3384&17.34$\pm$0.05&15.27$\pm$0.03&11.3\\
NGC3608&17.70$\pm$0.03&16.20$\pm$0.04&22.3\\
NGC4262&17.06$\pm$0.04&16.01$\pm$0.04&15.4\\
NGC4270&19.32$\pm$0.10&17.46$\pm$0.02&33.1\\
NGC4278&16.32$\pm$0.04&15.32$\pm$0.03&15.6\\
NGC4458&19.25$\pm$0.10&17.18$\pm$0.03&16.4\\
NGC4459&17.22$\pm$0.04&15.54$\pm$0.04&16.1\\
NGC4473&17.19$\pm$0.04&15.54$\pm$0.03&15.3\\
NGC4486&14.64$\pm$0.01&13.85$\pm$0.02&17.2\\
NGC4546&17.65$\pm$0.04&15.78$\pm$0.02&13.7\\
NGC4552&15.97$\pm$0.03&15.02$\pm$0.01&15.8\\
NGC4564&17.72$\pm$0.03&16.38$\pm$0.07&15.8\\
NGC4570&17.65$\pm$0.02&16.13$\pm$0.02&17.1\\
NGC4621&16.61$\pm$0.05&15.24$\pm$0.08&14.9\\
NGC4660&17.81$\pm$0.03&16.08$\pm$0.04&15.0\\
NGC5308&18.51$\pm$0.03&16.99$\pm$0.02&34.1\\
NGC5813&17.14$\pm$0.03&15.58$\pm$0.04&31.3\\
NGC5838&17.77$\pm$0.06&16.20$\pm$0.02&19.8\\
NGC5845&19.15$\pm$0.05&17.87$\pm$0.03&25.2\\
NGC5846&16.52$\pm$0.04&15.48$\pm$0.04&24.2\\
\enddata
\label{tab:dat}
\tablenotetext{a}{The quoted uncertainties do not include the zero point uncertainties, but a systematic zero point error would affect all measurements equally and so do not affect the conclusions presented here.}
\end{deluxetable}

We analyze ultraviolet imaging data for the 34 ETGs and the bulge of M~31
in the Conroy \& van Dokkum (2012) sample, where we obtained the most recently
pipeline-processed (GR6/7 release) {\sl GALEX} data from the MAST archive maintained by the Space Telescope Science Institute. For all 35 targets we have
imaging data of MIS-depth (one {\sl GALEX} orbit or more; mainly from the
NGS, GI, MIS, and DIS surveys) in the {\sl GALEX} NUV band and for 32 of
them, all except NGC~3414, NGC~4382, and NGC~4564, we also have similarly deep
FUV data. The first two of these three galaxies are
excluded from our analysis because only shallow ($\leq$100\,s-long exposures) FUV data from the All-sky Imaging
Survey (AIS) exist. In the
case of NGC~4564, the FUV total exposure time was 504 seconds, not
quite as long as for the rest of the FUV imaging data but enough for
the purpose of this study. We exclude the bulge of M 31 on the grounds that it is significantly different in nature than the other systems.

We follow the procedure
described by \cite{gdp} and \cite{lee}. In summary, the analysis steps
are (1) sky-background subtraction,
using elliptical annuli centered on the galaxy that match the
ellipticity and position angle (PA) of the galaxy and have major axes significantly larger
than the isophotal diameter, D25, in all cases (the same region is used for each of the two UV
bands), (2) interactive masking of foreground stars and background
galaxies following an automated detection of all red (FUV-NUV $>1$) point sources as potential
contaminants, 3) surface photometry within elliptical annuli with
fixed center, ellipticity and PA (those of the D25 ellipse) and (4)
calculation of the growth curve in both UV bands and the
derivation of the corresponding asymptotic magnitudes.
We use the FUV and NUV
asymptotic magnitudes as the best measure of the total UV emission of
our galaxies (Table \ref{tab:dat}). Distances are adopted from the on-line database NED. 

\begin{figure*}
\vskip -1in
\plotone{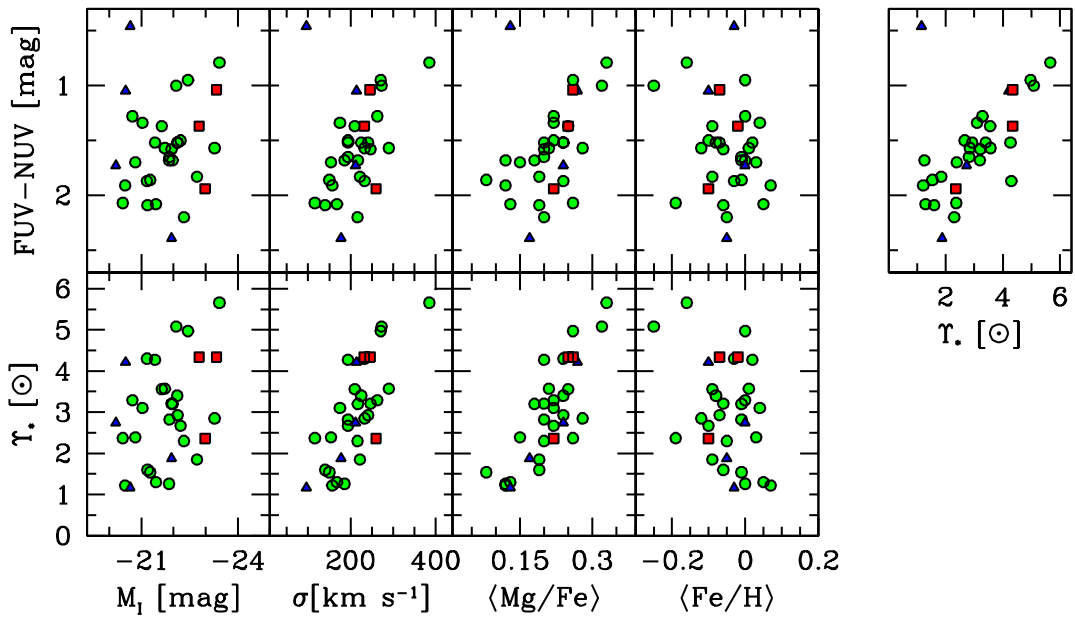}
\vskip -1in
\caption{Correlations among various key parameters. Symbols are as in Figure \ref{fig:UV} and described in text.}
\label{fig:comp}
\end{figure*}

\begin{deluxetable*}{lrrrrrr}
\tablecaption{Rank Correlation Coefficients and Probabilities They Arise Randomly\tablenotemark{a}}
\tablewidth{0pt}
\tablehead{
\colhead{}&
\colhead{FUV $-$ NUV}&
\colhead{M/L$_I$}&
\colhead{L$_I$}&
\colhead{$\sigma$}&
\colhead{$\langle Mg/Fe\rangle$}&
\colhead{$\langle Fe/H\rangle$} 
}
\startdata
FUV $-$ NUV&...&{\bf 6.3$\times$10$^{-8}$}&0.49&{\bf 2.0$\times$10$^{-3}$}&{\bf 3.1$\times$10$^{-5}$}&0.49\\
M/L$_I$&$-$0.78&...&0.23&{\bf 4.4$\times$10$^{-5}$}&{\bf 2.0$\times$10$^{-6}$}&0.25\\
L$_I$&0.12&$-$0.21&...&{\bf 8.9$\times$10$^{-3}$}&0.31&0.05\\
$\sigma$&$-$0.51&0.64&$-$0.44&...&{\bf 3.1$\times$10$^{-4}$}&0.03\\
$\langle Mg/Fe\rangle$&$-$0.65&0.72&$-$0.18&0.58&...&{\bf 4.3$\times$10$^{-5}$}\\
$\langle Fe/H\rangle$&0.12&$-$0.20&0.33&$-$0.38&$-$0.64&...\\
\enddata
\label{tab:rank}
\tablenotetext{a}{Rank Correlation Coefficients given in bottom half of matrix, probabilities in top half. Significant correlation highlighted in bold.}
\end{deluxetable*}

\section{Results}

We present the UV properties of the sample in Figure \ref{fig:UV} and Table \ref{tab:dat}. The top panel of the Figure shows the relation between $M_I$ and $M_{NUV}$, confirming that the UV flux is generally related to the properties of the galaxy as a whole and is not the product of UV ``froth" resulting from random low levels of recent star formation. We highlight the most striking outliers above the relationship  (UV-bright) using blue triangles. In the bottom panel we explore the nature of these outliers in color-magnitude space. These outliers are clearly UV-bright and UV-blue and may be examples of galaxies where recent star formation is affecting the total photometry. We also note three galaxies that are redder than the red sequence in this panel, which are highlighted using red squares. These objects might represent cases where reddening is somewhat more prevalent, or alternatively galaxies with different mean stellar populations due to a factor such as mean metallicity. The latter supposition is not supported by the available data.

The surprising result comes in the far right panel of Figure \ref{fig:comp}, where we show the relationship between UV color (FUV $-$ NUV) and $\Upsilon_*$. The relationship between FUV $-$ NUV and $\Upsilon_*$ is incredibly tight (Spearman $R = 0.78$, excluding one outlier --- see below), tighter even than that between $\Upsilon_*$ and the velocity dispersion, $\sigma_V$, found by \cite{conroy} and used to motivate the finding of a mass dependent IMF (see Table \ref{tab:rank} for a summary of the Spearman rank correlation coefficients in the lower half of the matrix and the probability that correlation coefficients that are at least as large as those measured arise by chance in the upper half). The one notable exception to the tight relationship between FUV $-$ NUV and $\Upsilon_*$ is NGC 2685, which is known as the ``Spindle" galaxy for its odd morphology \citep{sandage} and is modeled as a multiply-ringed polar-ring galaxy \citep{jozsa}. We have removed this one object from all of the correlation calculations in Table \ref{tab:rank}, but include it in all of the Figures. 
 
The galaxies we identified as the blue outliers are NGC 474, 2685, 4262, 4660. These are all well-known S0 galaxies, three of which show clear interaction signatures. We discussed NGC 2685 above and NGC 4262 has an extended UV ring \citep{bettoni}. NGC 474 is classified as peculiar on the basis of its  shells \citep{buta} and identified as having residual star formation by \cite{jeong}. The other three of our blue outliers are not  among the galaxies classified by \cite{jeong}, but three other galaxies from our sample (NGC 1023, NGC 2974, and NGC 4459) are identified as having residual star formation, although these are manifestly not large outliers in Figure \ref{fig:UV}. Of the four blue outliers that we identified, NGC 4660, is the most normal in appearance and its large disk component was only discovered through its kinematics \citep{rix}. We suspect that the unusual UV properties of these four are related to recent, and perhaps unusually strong, interactions rather than the S0 nature of the galaxies themselves because the three red outliers (NGC 524, NGC 4621, and NGC 5846) are also S0 galaxies and lack clear interaction signatures. Interestingly, ``blue" S0's are typically difficult to find in the types of environment where S0's are thought to be forming \citep[for example, see][]{just} and may represent a stage following the E+A or post-starburst phase \citep{yang}. Nevertheless, with the exception of NGC 2685, which is quite an unusual galaxy, the relationship between the UV-upturn color and $\Upsilon_*$ holds.

Of the other correlations explored in Table \ref{tab:rank} and Figure \ref{fig:comp}, only the ones with $\sigma_V$ and $\langle Mg/Fe \rangle$ are statistically significant, but they are both quantitatively somewhat weaker than that between FUV $-$ NUV and $\Upsilon_*$. 

We are concerned that the extremely strong correlation between FUV $-$ NUV and $\Upsilon_*$ results from a systematic error in the derivation of $\Upsilon_*$ caused either directly or indirectly by the population responsible for the UV-upturn. A direct influence is unlikely as any black body responsible for the UV flux would provide negligible flux in the I-band where the IMF sensitive lines lie. More plausible is that the presence of UV emitting stars is related to a ``distortion" of the AGB or RGB population, either affecting the line indices or lowering the I-band flux below expectations. The former, an effect on the line indices, cannot be responsible because the variation in $\Upsilon_*$ is also inferred from kinematic analyses \citep{cappellari,laesker}. If, however, a lower-than-expected I$-$band flux is the rule in galaxies with large UV-upturns, then both the line-index and kinematic method will yield higher $\Upsilon_*$ for those galaxies. In such a scenario, the measurement of $\Upsilon_*$ would not be in error, but the inference that variations in $\Upsilon_*$ are  related to a change in the IMF would be.

We now discuss two scenarios that attempt to explain the empirical findings with and without IMF variations. First, consider the situation where the UV upturn stars come directly from the population of AGB/RGB stars, which is to say that some process converts AGB/RGB stars into UV-upturn stars, and this phenomenon is not included in current spectral synthesis models. In such a scenario, proportionally more UV upturn stars in a certain galaxy means proportionally fewer AGB/RGB stars, and therefore a lower $I-$band luminosity.  A relationship in this sense is found in at least one model of EHB evolution \citep{buzzoni}, although it is unclear whether that particular model will work quantitatively for the UV upturn-$\Upsilon_*$ relationship we observe. Other possibilities, for example one in which chemical abundance plays a role through winds and mass transfer rates, would also help explain other observables such that  of larger UV-upturn populations in galaxies with higher Mg$_2$.
Second, consider the situation where low mass stars in multiple star systems play a key role in transforming more massive evolved stars into the EHB stars of the UV-upturn population. In such a scenario, having more low mass stars in a bottom-heavy stellar population will lead to a larger UV-upturn population in the galaxies with higher $\Upsilon_*$ --- again in qualitative agreement with the observations. We cannot, with the data discussed so far, distinguish between these two possibilities.

Photometric data in other passbands might help resolve the situation. In a scenario where 
the I-band luminosity is different than the model expectations --- whether the origin of that discrepancy is physical or observational --- we might expect to see significant deviations in colors involving the I-band. We obtain H-band data for much of the sample from 2MASS (collated from the NED database) and find no significant correlation between I-band and I$-$H. We conclude that variations in I-band luminosities alone are an unlikely origin for our findings.

Given 1) that the $\Upsilon_*$ correlation with $\sigma_V$ is observed when $\Upsilon_*$ is either measured spectroscopically or kinematically, 2) the lack of any signature of odd behavior in the I-band luminosities as reflected in I-H, 3) the lack of a dependence of the $\Upsilon_*$ of old (log(age)$>$9.8) Local Group stellar clusters with $\langle Mg/Fe \rangle$ \citep{cameron, colucci,z12,z13},  and 4) that the strongest correlation we find is between $\Upsilon_*$ and UV color, we conclude that the variation in $\Upsilon_*$ is real, and that it is directly correlated to UV color. We suggest that IMF variations are responsible for the differences in the EHB populations of ETGs. Nevertheless, we need to remain cognizant of the possibility that other physical effects beside IMF variations are responsible for the various correlations \citep[cf.][]{conroy}. The possibility that metallicity is somehow responsible for both the IMF and UV-upturn variations remains a real possibility. 

\section{Conclusion}

Using {\sl GALEX} data we have identified a strong correlation between UV color and the stellar mass to light ratio, $\Upsilon_*$ in early type galaxies (ETGs). This correlation is stronger, within the same sample, than those previously identified between $\Upsilon_*$ and either $\sigma_V$, metallicity, or alpha enhancement. 

We are faced with that task of explaining an interrelated set of significant correlations between $\sigma_V$, $\langle Mg/Fe \rangle$, $\Upsilon_*$, and UV color. 
It is inherently difficult to draw conclusions from a quantitative comparison among statistically significant correlations because the measured strength of the correlations depends on the magnitude of the observational uncertainties. We hypothesize that the variations in the IMF identified in previous investigations are driving the correlation between $\Upsilon_*$ and UV color, but the role of metallicity in both is still unknown. As is usually the case with empirical correlations, understanding their origin and determining whether a correlation implies causality is a more difficult task than the discovery of the correlation.

A final intriguing possibility is the use of the UV color as a tracer of IMF variations in galaxies without recent star formation, at least as a selection criteria with which to identify galaxies for further study. 

\begin{acknowledgments}

DZ acknowledges financial support from 
NASA ADAP NNX12AE27G and NSF AST-1311326, and NYU CCPP for its hospitality during long-term visits. 
The authors acknowledge the support from the FP7 Marie Curie Actions of the European Commission, via the Initial Training Network DAGAL under REA grant agreement PITN-GA-2011-289313.
This research has made use of the NASA/IPAC Extragalactic Database (NED), which is operated by the Jet Propulsion Laboratory, California Institute of Technology, under contract with NASA. \end{acknowledgments}


\begin{thebibliography}

\bibitem[Bastian et al.(2010)]{bastian}
Bastian, N., Covey, K.R., \& Meyer, M.R. 2010, {\sl ARA\&A}, 48, 339

\bibitem[Bettoni et al.(2010)]{bettoni}
Bettoni, D., Buson, L.M., \& Galleta, G. 2010, \aap, 519, 72

\bibitem[Boselli et al.(2005)]{boselli}
Boselli, A. et al. 2005, \apj, 629, L29

\bibitem[Buta et al.(2010)]{buta}
Buta, R.J. et al. 2010, \apjs, 190, 147

\bibitem[Brown et al.(1997)]{brown97}
Brown, T.M., Ferguson, H.C., Davidsen, A.F., \& Dorman, B. 1997, \apj, 482, 685

\bibitem[Brown et al.(2000)]{brown00}
Brown, T.M., Bowers, C.W., Kimble, R.A., Sweigart, A.V., \& Ferguson, H.C. 2000, \apj, 532, 308

\bibitem[Bruzual(1983)]{bruzual}
Bruzual, G. 1983, \apj, 273, 105

\bibitem[Bureau et al.(2011)]{bureau}
Bureau, M. et al. 2011, \mnras, 414, 1887

\bibitem[Burstein et al.(1988)]{burstein}
Burstein, D., Bertola, F., Buson, L.M., Faber, S.M., \& Lauer, T.R. 1988, \apj, 328, 440

\bibitem[Buzzoni et al.(2012)]{buzzoni12}
Buzzoni, A., Bertone, E., Carraro, G., \& Buson, L. 2012, \apj, 749, 35

\bibitem[Buzzoni \& Gonz\'alez-L\'opezlira(2008)]{buzzoni}
Buzzoni, A., \& Gonz\'alez-L\'opezlira, R.A. 2008, \apj, 666, 1007

\bibitem[Cameron(2009)]{cameron}
Cameron, S.A., 2009, Ph.D. Thesis, University of Michigan

\bibitem[Cappellari et al.(2012)]{cappellari}
Cappellari, M. et al. 2012, Nature, 484, 485

\bibitem[Ciardullo \& Demarque(1978)]{ciardullo}
Ciardullo, R.B., \& Demarque, P. 1978, in IAU Symposium 80: The HR Diagram, ed. A.G.D. Philip \& D.S. Haynes (Dordrecht:Reidel), 345

\bibitem[Code \& Welch(1979)]{code}
Code, A.D. \& Welch, G.A. 1979, ApJ, 228, 95

\bibitem[Colucci et al.(2012)]{colucci}
Colucci, J.E., Bernstein, R.A., Cameron, S.A., \& McWilliam, A. 2012, \apj, 746, 29

\bibitem[Conroy \& van Dokkum(2012)]{conroy}
Conroy, C., \& van Dokkum, P. G. 2012, \apj, 760, 71

\bibitem[Faber(1983)]{faber}
Faber, S.M., 1983, Highlights Astr. 6, 165

\bibitem[Ferguson et al.(1991)]{ferguson}
Ferguson, H.C., et al. 1991, \apjl, 382, 69

\bibitem[Ferreras et al.(2013)]{ferreras}
Ferreras, I., La Barbera, F., de la Rosa, I.G., Vazdekis, A., de Carvalho,  R.R., Falc\'on-Barroso, J. \& Ricciardelli, E. 2013, \mnras, 429, 15L

\bibitem[Gil de Paz et al.(2007)]{gdp}
Gil de Paz, A., et al. 2007, \apjs, 173, 185

\bibitem[Greggio \& Renzini(1983)]{greggio}
Greggio, L., \& Renzini, A. 1983, \aap,118, 217

\bibitem[Gunn et al.(1981)]{gunn}
Gunn, J.E., Stryker, L.L., \& Tinsley, B.M. 1981, \apj, 249, 48

\bibitem[Han et al.(2007)]{han}
Han, Z., Podsiadlowski, Ph. \& Lynas-Gray, A.E. 2007, \mnras, 380, 1098

\bibitem[Jeong et al.(2012)]{jeong}
Jeong, H. et al. 2012, \mnras, 423, 1921

\bibitem[J\'ozsa et al.(2008)]{jozsa}
J\'ozsa, G.I.G., Osterloo, T.A., Morganti, R., Klein, U., \& Erben, T. 2008, \aap, 494,489

\bibitem[Just et al.(2011)]{just}
Just, D., Zaritsky, D. Tran, K.-V. H., Gonzalez, A. H., Kautsch, S. J., \& Moustakas, J. 2011, \apj, 740, 54

\bibitem[Kalirai et al.(2012)]{kalirai}
Kalirai, J.S. et al. 2012, arXiv1212.1159

\bibitem[Kannappan et al.(2009)]{kannappan}
Kannappan, S.J., Guie, J.M., \& Baker, A.J. 2009, \aj, 138, 579

\bibitem[Kaviraj et al.(2007)]{kav}
Kaviraj, S. et al. 2007, \apjs, 173, 619

\bibitem[L\"asker et al.(2013)]{laesker}
L\"asker, R., van den Bosch, R.C.E., van de Ven, G., Ferreras, I., La Barbera, F., Vazdekis, A., \& Falc\'on-
Barroso, J. 2013, \mnras, in press

\bibitem[Lee et al.(2011)]{lee}
Lee, J.C., et al. 2011, \apjs, 192, 6

\bibitem[Lisker \& Han(2008)]{lisker}
Lisker, T., \& Han, Z. 2008, \apj, 680, 1042

\bibitem[Martin et al.(2005)]{martin}
Martin, D.C., et al. 2005, \apjl, 619, L1


\bibitem[Mu\~noz-Mateos, et al.(2013)]{munoz}
Mu\~noz-Mateos, J.C., et al. 2013, in prep.

\bibitem[Peacock, et al.(2011)]{peacock}
Peacock, M.B., Maccarone, T.J., Dieball, A., \& Knigge, C. 2011, \mnras, 411, 487

\bibitem[Rix \& White(1990)]{rix}
Rix, H.-W., \& White, S.D.M. 1990, \apj, 362, 52

\bibitem[Rocca-Volmerange \& Guiderdoni(1987)]{rv}
Rocca-Volmerange, B. \& Guiderdoni, B. 1987, \aap, 175, 15

\bibitem[Rose \& Tinsley(1974)]{rose}
Rose, W.B. \& Tinsley, B.M. 1974, \apj, 190, 243

\bibitem[Salim et al.(2012)]{salim}
Salim, S., Fang, J.J., Rich, R.M., Faber, S.M., \& Thilker, D.A. 2012, \apj, 755, 105

\bibitem[Sandage(1961)]{sandage}
Sandage, A. 1961, {\sl The Hubble Atlas of Galaxies} (Washington: Carnegie Institute of Washington)

\bibitem[Schawinski et al.(2007)]{scha}
Schawinski, K., et al. 2007, \apjs, 173, 512

\bibitem[Spiniello et al.(2012)]{spiniello}
Spiniello, C., Trager, S.C., Koopmans, L.V.E., \& Chen, Y.P. 2012, \apj, 753, 32

\bibitem[Strader et al.(2011)]{strader}
Strader, J., Caldwell, N., \& Seth, A.C. 2011, \aj, 142, 8

\bibitem[van Dokkum \& Conroy(2010)]{vandokkum}
van Dokkum, P.G. \& Conroy, C. 2010, Nature, 468, 940

\bibitem[Werner et al.(2004)]{werner}
Werner, M.W., et al. 2004, \apjs, 154, 1

\bibitem[Yang et al.(2008)]{yang}
Yang, Y., Zabludoff, A.I., Zaritsky, D., \& Mihos, J.C. 2008, \apj, 688, 945

\bibitem[Yi et al.(2005)]{yi}
Yi, S., et al. 2005, \apj, 619, L111

\bibitem[Zaritsky et al.(2012)]{z12}
Zaritsky, D., Colucci, J.E., Pessev, P.M., Bernstein, R.A., \& Chandar, R., 2012, \apj, 761, 93

\bibitem[Zaritsky et al.(2013)]{z13}
Zaritsky, D., Colucci, J.E., Pessev, P.M., Bernstein, R.A., \& Chandar, R., 2013, \apj, 770, 121

\end{thebibliography}
\end{document}